\documentclass[aps,prx,twocolumn,amsmath,amssymb]{revtex4-2}
\usepackage{graphicx}
\usepackage[colorlinks=true,citecolor=blue,linkcolor=blue,urlcolor=blue]{hyperref}
\usepackage{amsmath,amssymb,bm,physics}
\usepackage[colorlinks=true,citecolor=blue,linkcolor=blue,urlcolor=blue]{hyperref}
\usepackage{comment}
\usepackage[caption=false]{subfig}

\begin{document}
\title{Kitaev chain in synthetic dimension with cavity-controlled Majorana modes}

\author{Adel Ali}
\affiliation{Department of Physics and Astronomy, Texas A\&M University, College Station, TX, 77843 USA}
\author{Alexey Belyanin}
\affiliation{Department of Physics and Astronomy, Texas A\&M University, College Station, TX, 77843 USA}

\date{\today}

\begin{abstract}

We introduce a tunable synthetic-dimension platform for realizing
Kitaev-chain physics with high degree of control over Majorana zero modes.  It is based on a generic Landau-quantized two dimensional electron system coupled to the magnetic flux of a superconducting LC circuit. The structured vector potential of a superconducting LC inductor induces attractive interactions between electron angular-momentum states at the lowest Landau level. These states serve as a synthetic dimension for the coveted fermionic Kitaev chain, with Majorana zero modes existing at the
boundaries of the angular-momentum lattice. The crucial advantage of this proposal is the possibility of a robust, nonlocal readout and control of the Majorana states by a LC resonator. The platform relies on mature circuit QED and semiconductor technologies and provides a promising pathway to topological quantum computing.

\end{abstract}

\maketitle

\section{Introduction}

The search for controllable Majorana zero modes has become a central
theme in the study of topological quantum matter.  The minimal theoretical
paradigm is Kitaev's spinless \(p\)-wave chain, in which a one-dimensional
superconductor enters a class-D topological phase supporting exponentially
localized Majorana boundary modes \cite{Kitaev2001}.  More broadly,
Majorana zero modes are attractive for quantum information because the
fermion parity stored by spatially separated Majoranas is intrinsically
nonlocal, and braiding or measurement-only protocols can implement
fault-tolerant operations protected against local perturbations
\cite{Nayak2008,Alicea2012,Aasen2016}.

Physical realizations of Kitaev-chain physics have been pursued in several
platforms.  The most developed route uses semiconductor nanowires with
strong spin-orbit coupling, Zeeman splitting, and proximity-induced
\(s\)-wave superconductivity, which together emulate an effective spinless
\(p\)-wave superconductor \cite{Lutchyn2010,Oreg2010}.  Related proposals
use topological-insulator surfaces or Josephson structures to obtain
Majorana modes from conventional superconductors through spin-momentum
locking and proximity pairing \cite{FuKane2008}.  These platforms have led
to extensive experimental progress, including zero-bias anomalies and
hybrid nanowire devices suggestive of Majorana physics \cite{Mourik2012}.
Nevertheless, unambiguous realization and manipulation remain challenging.
The required conditions combine a hard induced superconducting gap,
controlled chemical potential, sufficiently large spin-orbit coupling,
strong but not destructive magnetic fields, low disorder, suppressed
quasiparticle poisoning, and reliable nonlocal readout of the Majorana
pair.  Moreover, braiding in one-dimensional wire implementations requires
networks, T-junctions, or measurement protocols rather than a single
isolated chain \cite{Alicea2011,Aasen2016}. More recently, artificial Kitaev chains based on quantum-dot superconductor
arrays have realized minimal two- and three-site versions of the model,
highlighting both the scalability challenge of real-space implementations \cite{Dvir2023,Bordin2025}. These challenges motivate
alternative routes in which the ingredients of the Kitaev chain are not
provided by real-space hopping and proximity pairing alone, but are instead
engineered in a more programmable Hilbert space.

Synthetic dimensions provide precisely such a route.  In a synthetic
dimension, a discrete set of internal, modal, frequency, orbital, or
momentum states is reinterpreted as the sites of an artificial lattice
\cite{Boada2012,Celi2014,OzawaPrice2019}.  Couplings between these states
play the role of hopping matrix elements, while phases of the couplings
generate synthetic gauge fields.  This idea has enabled topological
lattice models in systems whose physical dimension is lower than the
effective dimension of the Hamiltonian.  Momentum-space lattices in cold
atoms provide a particularly relevant precedent
\cite{Meier2016,An2018}.  Photonic frequency and orbital-angular-momentum
modes provide another important setting in which modal indices are used as
synthetic coordinates \cite{Yuan2018}.  These developments demonstrate
that a synthetic lattice need not correspond to a physical array of sites;
it may instead be encoded in a controllable basis of quantum states.

In parallel, a broader research area of ``cavity quantum materials'' has emerged, where the electromagnetic environment is an active mediator capable of reshaping many-body
interactions \cite{SchlawinKennesSentef2022}.  In particular, it has been
shown that structured cavity vacuum fluctuations can mediate
effective electron-electron interactions and induce superconducting
pairing, including pairing channels generated by current-current
interactions in two-dimensional electronic systems \cite{Schlawin2019,
Chakraborty2021}.  Circuit-QED architectures provide a potentially more programmable setting for this idea: the mode profile,
impedance, frequency, and inductive participation of a superconducting LC
resonator can be engineered, allowing the effective
interaction kernel to be shaped by circuit geometry rather than inherited
from a fixed optical cavity mode \cite{Blais2021,Krantz2019,Ali2026-yg}. 


\begin{figure}
    \centering
    \includegraphics[width=0.99\linewidth]{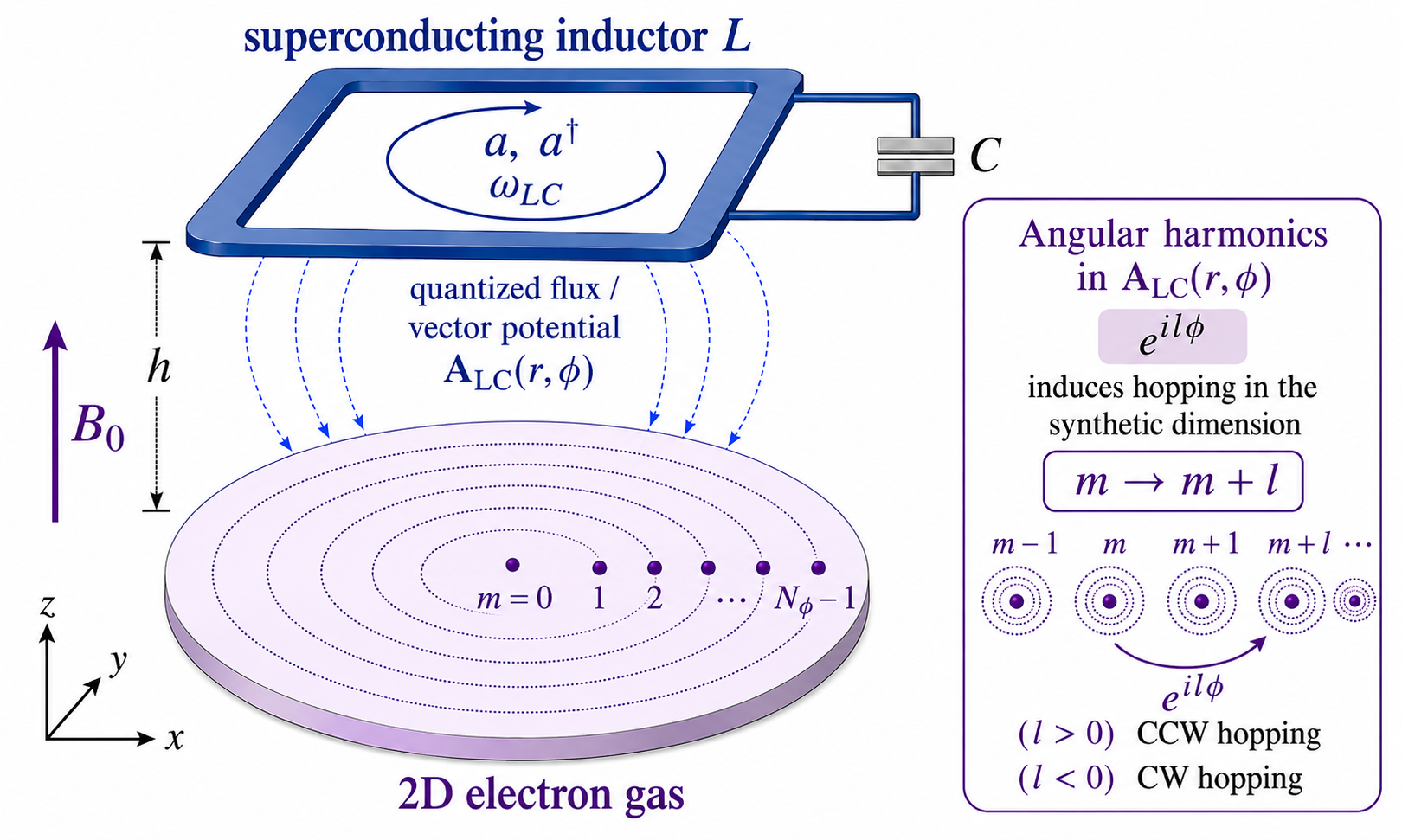}
    \caption{Schematic of a disk-shaped 2DEG sample placed in a strong static transverse magnetic field \(\mathbf B_0=B_0\hat{\mathbf z}\)  and proximally coupled to the inductive element of a superconducting \(LC\) circuit. Angular harmonics of the circuit field, \(\mathbf A_{\rm LC}(r,\phi)\sim e^{i\ell\phi}\), obey the selection rule \(\Delta m=\ell\), thereby inducing hopping \(m\rightarrow m+\ell\) along the synthetic angular-momentum dimension. A loop geometry dominated by \(\ell=0\) and \(\ell=1\) harmonics produces the nearest-neighbor synthetic hopping needed for a Kitaev-chain-like reduction, while higher harmonics generate longer-range corrections. 
    \label{fig:qh_lc_setup}}
\end{figure}

Here we bring these developments together to propose a new synthetic route to the coveted fermionic Kitaev chain and associated Majorana physics, based on a conceptually simple system:  a generic Landau-quantized two dimensional electron gas (2DEG), such as a semiconductor quantum well or graphene, coupled to a superconducting LC circuit.  Other than the underlying mature technologies, its crucial advantage is naturally long chains and the possibility of a robust, nonlocal microwave readout and control of the Majorana states. 

A simple conceptual example is sketched in Fig.~1. Electrons in a 2DEG sample are at the lowest Landau level (LLL) in a strong perpendicular magnetic field and coupled to a quantized vector potential supported by an LC loop, which is fabricated on a dielectric spacer of nm-scale thickness $h$ (e.g., a quantum well barrier, or an hBN layer in the case of graphene).

In the symmetric gauge, the LLL
orbitals are labeled by the guiding-center angular momentum
\(m=0,1,\ldots,N_\phi-1\), with orbital radius
\(R_m\simeq\sqrt{2m}\ell_B\).  The index \(m\) therefore forms a natural
one-dimensional synthetic coordinate with open boundaries.
Although we use symmetric-gauge orbitals to represent the LLL states, the
synthetic coordinate is not a gauge artifact.  The physical system is a
finite-size 2DEG sample with a geometry set by the lateral boundaries (or lateral confinement potential) and by the LC loop geometry.  For a disk-shaped sample with a specified center, the LLL
guiding-center angular momentum labels the radial position of the orbitals. 
\footnote{A different gauge would represent the same
projected operators in a different basis, but it would not change the
spectrum or the existence of topological domain walls.  The compact
nearest-neighbor form of the synthetic chain is therefore a consequence of
the rotational structure of the physical sample and circuit field, not of
an arbitrary gauge choice.}

Unlike cold-atom momentum-space
lattices, where hopping is generated by optical momentum transfer, here
hopping in the synthetic \(m\)-direction is generated by angular harmonics
of a quantized magnetostatic vector potential.  The coupling is purely dispersive as we assume that the LC frequency is much higher than the energies of electron $m$-states, so that the quantized superconducting flux mode remains in the vacuum ground state.  Moreover, the same LC
mode that engineers the synthetic hopping and interaction may also provide
a dispersive readout channel.  Therefore, flux-engineered Landau-level synthetic dimensions
offer a new route to programmable topological superconductivity beyond
real-space nanowire architectures.



\section{Description of the platform} 

We consider a spin-polarized two-dimensional electron gas in a strong
classical perpendicular magnetic field \(B_0\hat{\bm z}\), coupled to a
single superconducting LC mode; see Fig.~1. The classical
bias vector potential and the quantized LC vector potential must enter the
minimal-coupling Hamiltonian before projection onto a Landau level.  For
electrons of charge \(-e\) and mass $m^*$, the total vector potential is written
\begin{equation}
    \bm A_{\rm tot}(\bm r)
    =
    \bm A_B(\bm r)
    +
    \hat X\,\bm A_{\rm LC}(\bm r),
    \qquad
    \hat X=\hat a+\hat a^\dagger ,
    \label{eq:Atot}
\end{equation}
where \(\bm A_B=(B_0 r/2)\hat{\bm\phi}\) in symmetric gauge.  The
continuum Hamiltonian is
\begin{align}
\hat H & = \int d^2r\, \hat\psi^\dagger(\bm r) \bigg[\frac{1}{2m^\ast} \left( -i\hbar\nabla + e\bm A_B + e\hat X\bm A_{\rm LC} \right)^2 \\
&+ V_{\rm conf}(\bm r) - \mu \bigg] \hat\psi(\bm r) + \hbar\omega_{\rm LC}\hat a^\dagger\hat a .
\label{eq:Hminimal}
\end{align}
Defining the mechanical momentum in the background field,
\begin{equation}
    \bm\Pi_B=-i\hbar\nabla+e\bm A_B ,
\end{equation}
and expanding Eq.~\eqref{eq:Hminimal} in powers of \(\hat X\), one obtains
\begin{equation}
    \hat H
    =
    \hat H_B
    +
    \hbar\omega_{\rm LC}\hat a^\dagger\hat a
    +
    \hat X\hat\Gamma
    +
    \hat X^2\hat D ,
    \label{eq:Hexpanded}
\end{equation}
with
\begin{equation}
    \hat H_B
    =
    \int d^2r\,
    \hat\psi^\dagger
    \left[
    \frac{\bm\Pi_B^2}{2m^\ast}
    +
    V_{\rm conf}
    -
    \mu
    \right]
    \hat\psi ,
\end{equation}
where 
\begin{equation}
    \hat\Gamma
    =
    \frac{e}{2m^\ast}
    \int d^2r\,
    \hat\psi^\dagger(\bm r)
    \left\{
    \bm A_{\rm LC}(\bm r),
    \bm\Pi_B
    \right\}
    \hat\psi(\bm r),
    \label{eq:Gamma}
\end{equation}
and
\begin{equation}
    \hat D
    =
    \frac{e^2}{2m^\ast}
    \int d^2r\,
    \hat n(\bm r)
    |\bm A_{\rm LC}(\bm r)|^2 .
    \label{eq:Dterm}
\end{equation}
Equation~\eqref{eq:Gamma} is the gauge-consistent linear LC vertex.  It
contains both the canonical paramagnetic-current term and the cross term
between the bias field \(B_0\) and the LC vector potential.  The diamagnetic term \(D\) given by Eq.~\eqref{eq:Dterm} is retained because it is required by minimal
coupling and controls the stability of the dispersive reduction.


\subsection{Landau-level projection} 

We now project onto the lowest Landau. In the symmetric gauge the
single-particle orbitals are
\begin{equation}
    \phi_m(r,\phi)
    =
    \frac{1}{\sqrt{2\pi\ell_B^2m!}}
    \left(
    \frac{r}{\sqrt{2}\ell_B}
    \right)^m
    e^{im\phi}
    e^{-r^2/4\ell_B^2},
    \label{eq:LLLorbitals}
\end{equation}
where \(\ell_B=\sqrt{\hbar/eB_0}\).  The projected field operator is
\begin{equation}
    \hat\psi(\bm r)
    =
    \sum_{m=0}^{N_\phi-1}
    \phi_m(\bm r)c_m .
       \nonumber 
\end{equation}
The LLL index \(m\) is therefore a one-dimensional synthetic coordinate.
The projected Hamiltonian has the form
\begin{equation}
    \hat H_{\rm LLL}
    =
    \hat H_{\rm el}
    +
    \hbar\omega_{\rm LC}\hat a^\dagger\hat a
    +
    \hat X\hat\Gamma_{\rm LLL}
    +
    \hat X^2\hat D_{\rm LLL},
    \label{eq:HLLL}
\end{equation}
where
\begin{equation}
    \hat H_{\rm el}
    =
    \sum_{m,n}h^{(0)}_{mn}c_m^\dagger c_n
       \nonumber 
\end{equation}
contains the confinement, disorder, and possible weak Landau-level
nonflatness.

The angular structure of the LC loop controls the range of hopping in the
synthetic \(m\)-direction.  We take the dominant LC vector potential in the
plane to be azimuthal,
\begin{equation}
    \bm A_{\rm LC}(r,\phi)
    =
    A_\phi(r)f(\phi)\hat{\bm\phi}.
       \nonumber 
\end{equation}
Expanding
\begin{equation}
    f(\phi)
    =
    \sum_{\ell=-\infty}^{\infty}
    f_\ell e^{i\ell\phi},
    \qquad
    f_{-\ell}=f_\ell^\ast ,
       \nonumber 
\end{equation}
the LLL angular integral gives the selection rule
\begin{equation}
    e^{i\ell\phi}
    \quad\Rightarrow\quad
    \Delta m=\ell .
    \nonumber 
\end{equation}
For example, an \(\ell=1\) harmonic generates nearest-neighbor matrix elements
\(m\leftrightarrow m+1\) in the Landau-level synthetic chain. 

Therefore, the geometry of the superconducting circuit directly controls the
range and phase of hopping in the Landau-level synthetic dimension.  In
particular, a nearly circular loop with a controlled dipolar asymmetry
contains dominant \(\ell=0\) and \(\ell=1\) harmonics, producing the
ingredients closest to a nearest-neighbor Kitaev chain. The angular profile that best realizes the synthetic chain is
\begin{equation}
    f(\phi)
    \simeq
    f_0
    +
    f_1\cos(\phi-\phi_0),
    \quad
    |f_{\ell\ge2}|\ll |f_1|\ll |f_0|.
       \nonumber 
\end{equation}

One of the geometries closest to this angular profile is a slightly off-center or single-lobed
loop. Keeping only these two harmonics, the projected LC vertex is
\begin{equation}
    \hat\Gamma_{\rm LLL}
    =
    \hat\Gamma_0+\hat\Gamma_1 ,
       \nonumber 
\end{equation}
with
\begin{equation}
    \hat\Gamma_0
    =
    \sum_m \eta_m c_m^\dagger c_m ,
       \nonumber 
\end{equation}
and
\begin{equation}
    \hat\Gamma_1
    =
    \sum_m
    \left(
    \gamma_m e^{-i\phi_0}c_{m+1}^\dagger c_m
    +
    \gamma_m^\ast e^{i\phi_0}c_m^\dagger c_{m+1}
    \right).
       \nonumber 
\end{equation}
The coefficients \(\eta_m\) and \(\gamma_m\) are radial LLL matrix
elements determined by \(A_\phi(r)\), the bias field, and the loop
geometry.  In the bulk of a large droplet one may approximate
\begin{equation}
    \eta_m\simeq\eta_0,\qquad \gamma_m\simeq \gamma_1 ,
       \nonumber 
\end{equation}
up to smooth edge corrections.

\subsection{Dispersive elimination of the LC mode.}

 In the off-resonant regime, the LC oscillator may be eliminated.  Replacing
\(\hat D_{\rm LLL}\) by its expectation value \(D^{(0)}\), one obtains
\begin{equation}
    \hat H_{\rm eff}
    =
    \hat H_{\rm el}
    -
    g_{\rm LC}\hat\Gamma_{\rm LLL}^2,
    \qquad
    g_{\rm LC}
    =
    \frac{1}{\hbar\omega_{\rm LC}+4D^{(0)}} .
    \label{eq:Heff}
\end{equation}
This form is the static limit of a Schrieffer-Wolff or oscillator
completion-of-the-square reduction.  Corrections are controlled by the
ratios of electronic energy scales to \(\hbar\omega_{\rm LC}\).

We now show how Eq.~\eqref{eq:Heff} produces the Kitaev chain.  In a fixed-particle-number sector, the nearly uniform
\(\ell=0\) piece acts as
\begin{equation}
    \hat\Gamma_0\simeq \eta_0 N_e .
    \nonumber 
\end{equation}
The cross term between \(\hat\Gamma_0\) and \(\hat\Gamma_1\) gives
\begin{align}
    -g_{\rm LC}
    \left(
    \hat\Gamma_0\hat\Gamma_1+\hat\Gamma_1\hat\Gamma_0
    \right)
    =
    -t_1
    \sum_m
    \left(
    e^{-i\phi_0}c_{m+1}^\dagger c_m \right.  \nonumber \\
   \left. +
    e^{i\phi_0}c_m^\dagger c_{m+1}
    \right),
    \label{eq:t1term}
\end{align}
where
\begin{equation}
    t_1
    =
    2g_{\rm LC}\eta_0\gamma_1 N_e .
    \label{eq:t1}
\end{equation}
The phase \(\phi_0\) is a Peierls phase in the synthetic chain and may be
removed by \(c_m\rightarrow e^{im\phi_0}c_m\) for an open chain.  Hence
the \((\ell=0)\times(\ell=1) \) cross term supplies the crucial nearest-neighbor
normal hopping required for a Kitaev chain.

The square of the \(\ell=1\) vertex produces subleading corrections.
For uniform \(\gamma_m=\gamma_1\), define
\begin{equation}
    T_1
    =
    \sum_m
    \left(
    c_{m+1}^\dagger c_m
    +
    c_m^\dagger c_{m+1}
    \right).
     \nonumber 
\end{equation}
Then
\begin{equation}
    T_1^2\big|_{\rm one-body}
    =
    2\sum_m n_m
    +
    \sum_m
    \left(
    c_{m+2}^\dagger c_m+\mathrm{h.c.}
    \right).
     \nonumber 
\end{equation}
Thus Eq.~\eqref{eq:Heff} also generates a chemical-potential shift and a
second-neighbor synthetic hopping,
\begin{equation}
    -t_2
    \sum_m
    \left(
    c_{m+2}^\dagger c_m+\mathrm{h.c.}
    \right),
    \qquad
    t_2\simeq g_{\rm LC}|\gamma_1|^2 .
    \label{eq:t2}
\end{equation}
The desired Kitaev limit is obtained when
\begin{equation}
    |t_2|\ll |t_1|,
     \nonumber 
\end{equation}
which is naturally favored by a dominant \(\ell=0\) component and a weak
dipolar distortion.

\subsection{Reduction to Kitaev Hamiltonian}

 The same interaction in Eq.~\eqref{eq:Heff} contains a genuine two-body
term.  For a general one-body vertex
\begin{equation}
    \hat\Gamma_{\rm LLL}=\sum_{m,n}\Gamma_{mn}c_m^\dagger c_n ,
     \nonumber 
\end{equation}
normal ordering gives
\begin{equation}
    \hat\Gamma_{\rm LLL}^2
    =
    \sum_{m,q}
    (\Gamma^2)_{mq}c_m^\dagger c_q
    -
    \sum_{m,n,p,q}
    \Gamma_{mn}\Gamma_{pq}
    c_m^\dagger c_p^\dagger c_n c_q .
     \nonumber 
\end{equation}
Therefore
\begin{equation}
    \hat H_{2{\rm b}}
    =
    g_{\rm LC}
    \sum_{m,n,p,q}
    \Gamma_{mn}\Gamma_{pq}
    c_m^\dagger c_p^\dagger c_n c_q .
    \label{eq:H2b}
\end{equation}
The leading superconducting instability is obtained by solving the
antisymmetrized Cooper-channel eigenvalue problem associated with
Eq.~\eqref{eq:H2b}.  If the leading attractive eigenmode is the nearest-neighbor odd channel, the natural order parameter is
\begin{equation}
    \Delta_m
    =
    U_{\rm eff}
    \langle c_m c_{m+1}\rangle ,
    \qquad
    \Delta_m\simeq\Delta ,
    \label{eq:DeltaDef}
\end{equation}
where \(U_{\rm eff}>0\) denotes the magnitude of the attractive eigenvalue.
This step is a controlled mean-field projection of the full interaction,
not an identity: competing particle-hole orders generated by
\(-g_{\rm LC}\hat\Gamma^2\) must be compared with the Cooper channel.

Under this Cooper-channel reduction, the mean-field pairing Hamiltonian is
\begin{equation}
    \hat H_\Delta
    =
    \sum_m
    \left(
    \Delta c_m^\dagger c_{m+1}^\dagger
    +
    \Delta^\ast c_{m+1}c_m
    \right).
    \label{eq:HDelta}
\end{equation}
Combining Eqs.~\eqref{eq:t1term}, \eqref{eq:t2}, and \eqref{eq:HDelta}
gives the extended Kitaev Hamiltonian in the angular-momentum space,
\begin{align} 
    \hat H_{\rm K}^{\rm ext}
   & =
    -\mu_{\rm eff}\sum_m c_m^\dagger c_m
    -
    t_1\sum_m
    \left(
    c_{m+1}^\dagger c_m+\mathrm{h.c.}
    \right)
    \nonumber\\
    &
    -
    t_2\sum_m
    \left(
    c_{m+2}^\dagger c_m+\mathrm{h.c.}
    \right)
    +
    \sum_m
    \left(
    \Delta c_m^\dagger c_{m+1}^\dagger+\mathrm{h.c.}
    \right)
    \label{eq:extendedKitaev}
\end{align}
The standard Kitaev chain is recovered in the limit \(t_2/t_1\rightarrow0\).
The open boundaries of this chain are the two ends of the Landau-level
orbital sequence, \(m=0\) and \(m=N_\phi-1\).

To obtain the bulk topological criterion for the existence of Majorana edge modes, we make a Fourier transform along the synthetic
dimension,
\begin{equation}
    c_m
    =
    \frac{1}{\sqrt{N_m}}
    \sum_q e^{iqm}c_q .
        \nonumber 
\end{equation}
Using the Nambu spinor
$\Psi_q=
    \begin{pmatrix}
    c_q\\
    c_{-q}^\dagger
    \end{pmatrix},$
one obtains
\begin{equation}
    \hat H_{\rm BdG}
    =
    \frac12
    \sum_q
    \Psi_q^\dagger
    \mathcal H_{\rm BdG}(q)
    \Psi_q ,
        \nonumber 
\end{equation}
with
\begin{equation}
    \mathcal H_{\rm BdG}(q)
    =
    \xi(q)\tau_z
    +
    2|\Delta|\sin q\,\tau_y ,
    \label{eq:BdGq}
\end{equation}
and
\begin{equation}
    \xi(q)
    =
    -\mu_{\rm eff}
    -
    2t_1\cos q
    -
    2t_2\cos 2q .
    \label{eq:xiq}
\end{equation}
The quasiparticle spectrum is
\begin{equation}
    E(q)
    =
    \sqrt{
    \xi^2(q)
    +
    4|\Delta|^2\sin^2 q
    } .
        \nonumber 
\end{equation}
The class-D invariant changes only when the gap closes at
\(q=0\) or \(q=\pi\).  Hence the topological criterion is
\begin{equation}
    \xi(0)\xi(\pi)<0 ,
    \label{eq:topcriterion}
\end{equation}
or, explicitly, 
\begin{equation}
    \left(
    -\mu_{\rm eff}-2t_1-2t_2
    \right)
    \left(
    -\mu_{\rm eff}+2t_1-2t_2
    \right)
    <0 .
    \label{eq:topcriterionexplicit}
\end{equation}
For \(t_2\ll t_1\), this reduces to the usual Kitaev condition
\begin{equation}
    |\mu_{\rm eff}|<2|t_1| .
        \nonumber 
\end{equation}


\subsection{The readout scheme}

 A natural advantage of the proposed Landau-level synthetic Kitaev platform is that a weak LC readout mode can measure the Majorana parity dispersively.  Let \(a_r\) denote the readout resonator annihilation operator, \(\omega_r\) its bare frequency, and \(\gamma_i,\gamma_j\) the two Majorana zero modes whose fermion parity is
\[
P_{ij}=i\gamma_i\gamma_j=\pm1 .
\]
If the readout field weakly modulates a control parameter \(\lambda\) of the topological annulus, such as the synthetic chemical potential, hopping, pairing amplitude, or domain-wall position, then energy splitting between the two parity states $\varepsilon_{ij}$ becomes parity dependent,
\[
H_{\rm M}=\frac{\varepsilon_{ij}(\lambda)}{2}P_{ij}.
\]
Expanding \(\lambda\rightarrow \lambda+\lambda_{\rm zpf}(a_r+a_r^\dagger)\) gives a dispersive cavity shift
\[
H_{\rm read}
=
\hbar\left(\omega_r+\chi_{ij}P_{ij}\right)a_r^\dagger a_r,
\qquad
\chi_{ij}
\simeq
\frac{\lambda_{\rm zpf}^2}{2\hbar}
\frac{\partial^2\varepsilon_{ij}}{\partial\lambda^2}.
\]
Thus the reflected microwave phase measures \(P_{ij}=i\gamma_i\gamma_j\) without requiring a local tunnel probe attached to either Majorana.  This is advantageous in the present system because the zero modes are synthetic-boundary modes of the angular-momentum chain, corresponding to radially separated structures in the quantum Hall droplet.  Standard nanowire readout typically relies on quantum dots, tunnel contacts, Coulomb-blockaded islands, or interferometric probes near individual wire ends, whereas here the LC field naturally couples to an extended orbital operator and can provide a nonlocal, parity-sensitive, and potentially QND readout, provided \(\hbar\omega_r\), the resonator linewidth \(\hbar\kappa\), and \(k_BT\) remain below the bulk quasiparticle gap.



\section{Energy scales and electrostatic stability}

 We estimate the relevant scales for a GaAs quantum well  with comparable sample and LC loop radii, $R_s =R_{\rm loop}=4~\mu{\rm m}$, $B_0=1$ T,
$m^\ast=0.067m_e$, and 
$\epsilon_r=12.9$. 
The magnetic length and LLL degeneracy are
\[
\ell_B=25.7~{\rm nm},
\; 
N_\phi=\frac{B_0\pi R_{\rm QH}^2}{\Phi_0}\simeq1.2\times10^4;
\;
\Phi_0=\frac{h}{e}.
\]
We take the LC-loop height to be \(h_{\rm LC}=2~{\rm nm}\), and assume the metallic (or highly doped) screening gate at a distance $d$ below the 2DEG, typically tens of nm.

For an LC mode with \(f_{\rm LC}=100~{\rm GHz}\) and \(L=1~{\rm pH}\),
\[
\hbar\omega_{\rm LC}=414~\mu{\rm eV},\qquad
I_{\rm zpf}=\sqrt{\frac{\hbar\omega_{\rm LC}}{2L}}\simeq5.8~\mu{\rm A},
\]
giving
\[
B_{\rm zpf}\sim\frac{\mu_0I_{\rm zpf}}{2R_{\rm loop}}\simeq0.9~\mu{\rm T},
\qquad
\Phi_{\rm zpf}/\Phi_0\sim1.1\times10^{-2}.
\]
The corresponding edge-orbital matrix element is estimated as
\(\gamma_{\rm edge}\sim9.5~\mu{\rm eV}\); see the Supplemental Material.  For a strongly dipolar loop
profile,
\[
\epsilon_1\equiv |\gamma_1/\gamma_0|\simeq0.3,
\]
the \(\ell=1\) matrix element is $\gamma_1\sim2.9~\mu{\rm eV}$, and 
\[
V_{\rm LC}^{(1)}
\sim\frac{\gamma_1^2}{\hbar\omega_{\rm LC}}
\simeq2.0\times10^{-2}~\mu{\rm eV}.
\]
For an active Kitaev window of \(N_{\rm act}\sim10^3\) partially filled LLL
orbitals, the collective odd-channel eigenvalue is therefore
\[
\lambda_{\rm LC}^{\rm odd}
\sim N_{\rm act}V_{\rm LC}^{(1)}
\sim20~\mu{\rm eV}.
\]
We thus estimate an induced pairing scale $\Delta_1\sim10-20~\mu{\rm eV}$, 
and hence a topological gap
\[
E_{\rm gap}
=
\min_q\sqrt{\xi^2(q)+4|\Delta_1|^2\sin^2q}
\simeq2|\Delta_1|
\sim20-40~\mu{\rm eV},
\]
corresponding to
\[
T_{\rm gap}\sim0.2-0.5~{\rm K}.
\]

The relevant electrostatic energy is the LLL-projected adjacent-bond cost
in the same pairing channel,
\[
U_m^C=\langle m,m+1|V_C|m,m+1\rangle_A ,
\]
not the full Coulomb energy.  A metallic gate at distance \(d\) gives the
screened interaction
\[
V_C(q)=\frac{e^2}{2\epsilon_0\epsilon_r q}\left(1-e^{-2qd}\right).
\]
Taking \(d=30~{\rm nm}\) as an example, the central value is large,
\[
U_0^C\simeq0.67~{\rm meV},
\]
so the pairing near \(m=0\) is strongly suppressed.  However, the adjacent-bond
cost falls rapidly at large angular momenta because the orbitals become
extended annuli.  In an annulus with $R_{\rm in}=2~\mu{\rm m}$, $R_{\rm out}=4~\mu{\rm m}$, 
the active \(m\)-window is
\[
m_{\rm in/out}\simeq \frac{R_{\rm in/out}^2}{2\ell_B^2},
\; {\rm or} \;
m_{\rm in}\simeq3.0\times10^3,\;
m_{\rm out}\simeq1.2\times10^4 .
\]
Within this window, the screened adjacent-bond repulsion is only
\[
U^C_{m_{\rm in}}\sim2\times10^{-2}~\mu{\rm eV},
\quad
U^C_{m_{\rm out}}\sim3\times10^{-3}~\mu{\rm eV},
\]
which is negligible compared with the estimated gap
\(E_{\rm gap}\sim20-40~\mu{\rm eV}\).  Thus a gate-screened or annular
geometry as in Fig.~2(d) removes the small-\(m\) orbitals where electrostatic repulsion is
largest and places the active synthetic Kitaev chain in a regime where
Coulomb repulsion does not destroy the pairing instability. 

Finally, the dispersive condition is controlled by
\[
E_{\rm gap}/\hbar\omega_{\rm LC}\sim0.05-0.10 .
\]
At \(B_0=1~{\rm T}\), the GaAs cyclotron energy is
\[
\hbar\omega_c=\frac{\hbar eB_0}{m^\ast}\simeq1.73~{\rm meV},
\]
so the \(100~{\rm GHz}\) LC mode remains below the inter-Landau-level
cyclotron transition.  The chosen frequency is therefore high enough to
justify integrating out the LC mode on the Kitaev scale, while remaining
detuned from cyclotron Landau-polariton mixing.

As shown in the Supplemental Material, the estimates are even more favorable for InSb quantum wells due to about 5 times lower effective electron mass and stronger dielectric screening. In particular, one obtains about 5-10 higher topological gap and the gap temperature.


\section{Geometry and fabrication considerations}

\begin{figure}
    \centering
    \includegraphics[width=0.99\linewidth]{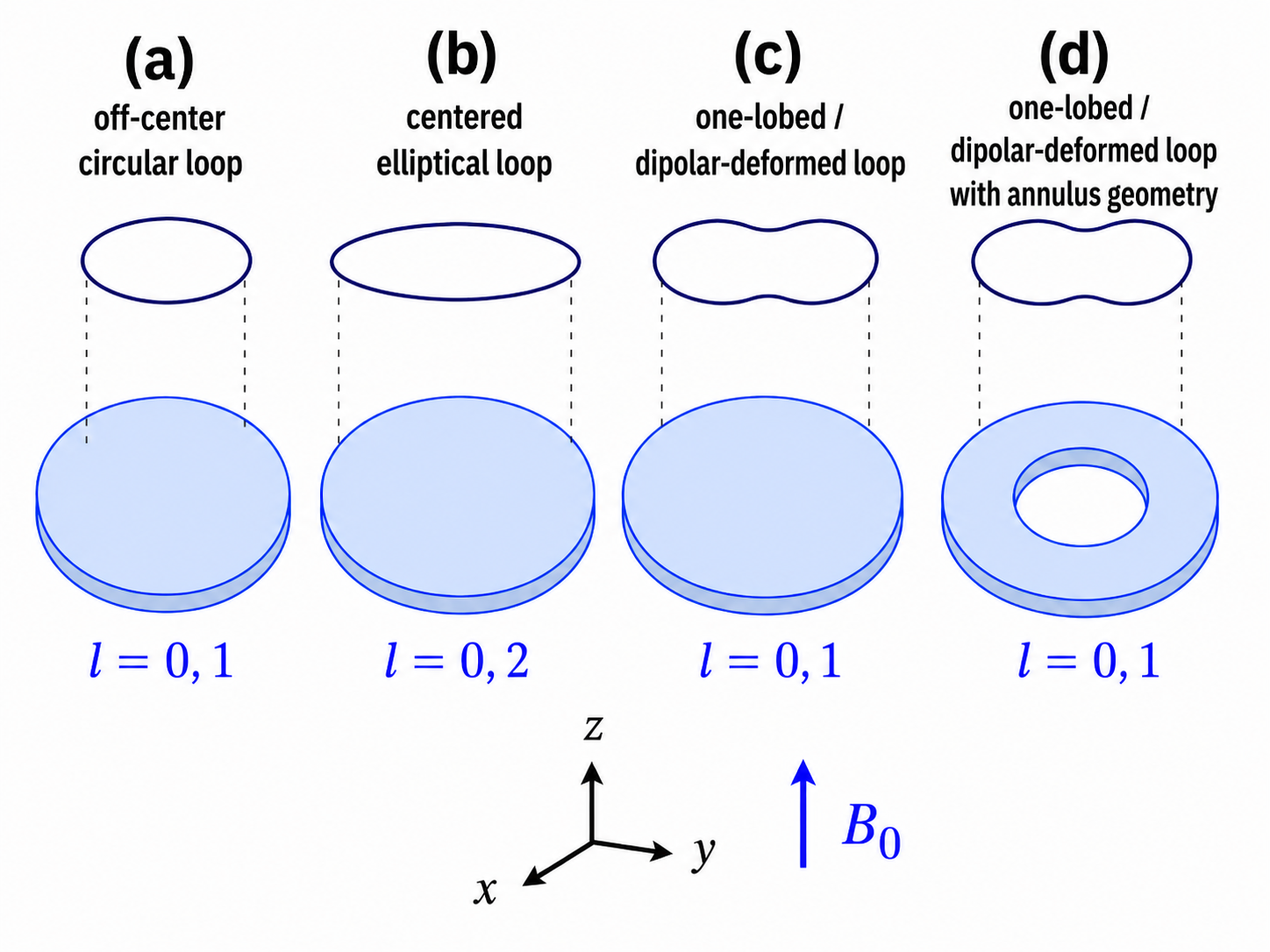}
    \caption{
Representative loop geometries for the inductive element of the superconducting \(LC\) circuit, shown above a 2DEG sample in a strong perpendicular magnetic field \(B_0\hat{\mathbf z}\), with dominant angular harmonics $\ell$ indicated.
The geometries dominated by \(\ell=0\) and \(\ell=1\) provide the closest route to a nearest-neighbor Kitaev-chain Hamiltonian, while even-harmonic geometries generate longer-range or two-sublattice couplings.
}
\label{fig:loop_geometries}
\end{figure}

Beyond the primary instability discussed above, the same architecture provides a direct route to engineering interactions in the synthetic orbital dimension.  The LC loop defines a quantized current
\begin{equation}
    \hat I
    =
    I_{\rm zpf}
    \left(\hat a+\hat a^\dagger\right),
    \qquad
    I_{\rm zpf}
    =
    \sqrt{\frac{\hbar\omega_{\rm LC}}{2L}},
    \label{eq:Izpf}
\end{equation}
where \(L\) is the loop inductance.  For a lithographically defined wire contour \(\mathcal C\), the vector potential per unit current is
\begin{equation}
    \mathbf a_{\mathcal C}(\mathbf r)
    =
    \frac{\mu_0}{4\pi}
    \oint_{\mathcal C}
    \frac{d\mathbf R}{|\mathbf r-\mathbf R|}.
    \label{eq:aC}
\end{equation}

Consequently, the paramagnetic light--matter coupling takes the compact form
\begin{equation}
    \hat H_{\rm para}
    =
    -\int d^2r\,
    \hat{\mathbf j}_p(\mathbf r)
    \cdot
    \mathbf a_{\mathcal C}(\mathbf r)I_{\rm zpf}
    \left(\hat a+\hat a^\dagger\right)
    =
    -(\hat a+\hat a^\dagger)\,\hat J_{\mathcal C}.
    \nonumber
    \label{eq:Hpara}
\end{equation}
This expression makes it explicit that the matter operator $\hat J_{\mathcal C}$ inherits the spatial and angular structure of the fabricated loop.  Thus, by shaping \(\mathcal C\), one controls the harmonic content of \(\mathbf a_{\mathcal C}\) and hence the selection rules and matrix elements between different synthetic-dimension states, such as angular-momentum orbitals in a Landau-quantized 2DEG sample.  The loop geometry therefore acts as a lithographically programmable interaction kernel rather than merely as a passive mode.
The geometry determines which synthetic hoppings are generated, as sketched in Fig.~2.  A centered
circular loop gives only an \(\ell=0\) vertex and therefore no nearest-neighbor synthetic hopping.  A centered ellipse has inversion symmetry,
\(f(\phi+\pi)=f(\phi)\), and therefore contains mainly even harmonics,
\begin{equation}
    f_{\rm ellipse}(\phi)
    =
    f_0+f_2\cos2\phi+f_4\cos4\phi+\cdots .
    \nonumber 
\end{equation}
It naturally generates \(m\rightarrow m\pm2\) and higher even-range
processes, producing two-sublattice extended Kitaev physics rather than the
single-chain Kitaev limit.

The optimal geometry for the present proposal is instead a nearly circular
loop with a controlled dipolar asymmetry,
\begin{equation}
    R_{\rm loop}(\phi)
    =
    R_0
    \left[
    1+\epsilon\cos(\phi-\phi_0)
    \right],
    \quad
    \epsilon < 1 ,
    \label{eq:limaçon}
\end{equation}
or equivalently a circular loop displaced slightly from the center of the
electron droplet.  Such a geometry produces
\begin{equation}
    f(\phi)
    =
    f_0
    +
    f_1\cos(\phi-\phi_0)
    +
    O(\epsilon^2),
        \nonumber 
\end{equation}
so that the unwanted higher harmonics are parametrically suppressed.  The
design rule for approaching the Kitaev limit is therefore
\begin{equation}
    |f_0 f_1|\ \text{large},
    \qquad
    |f_{\ell\ge2}|\ \text{small}.
        \nonumber 
\end{equation}
This can be implemented using standard planar superconducting-circuit
fabrication: a lithographic LC loop, SQUID loop, or high-kinetic-inductance
superconducting trace can be patterned with a small displacement, a smooth
single-lobed deformation, or a local side bulge.  The smooth one-lobed
design is preferred over sharp notches because it enhances the desired
\(\ell=1\) harmonic without introducing large high-\(\ell\) Fourier
components.


\section{Conclusions}

 We have shown a circuit-QED route to synthetic Kitaev-chain physics in a
Landau-quantized two-dimensional electron system.  The natural
one-dimensional synthetic lattice in LLL, with the structured magnetostatic vector
potential of a superconducting LC resonator generates controlled couplings
and effective attractive interactions within this lattice.  For suitable
angular harmonics of the LC field, the resulting projected Hamiltonian maps
onto an effective fermionic Kitaev chain. This provides a route to long effective
chains. In addition, electrostatic confinement
can tune the occupied angular-momentum window, while circuit parameters such
as the resonator frequency, impedance, and inductive participation control
the strength and structure of the mediated interaction. The same 
mode that engineers the interaction or a different weakly coupled mode offers a natural nonlocal channel
for dispersive readout and control of the Majorana degrees of freedom.

These features make our proposal a promising architecture for programmable
topological superconductivity.  Future work will address
measurement protocols, and synthetic-dimension braiding schemes based on
time-dependent gates and cavity drives, including measurement-only protocols
\cite{Aasen2016}.  

Recent theoretical and experimental developments in Quantum Hall systems coupled to cavity resonators \cite{Hagenmuller2010,
Scalari2012,Maissen2014,Appugliese2022,Enkner2024,Enkner2025,Ciuti2021,Rokaj2022} 
show that cavity vacuum fields can generate effective
nonlocal electronic processes in Landau-quantized matter. These experiments share similar ingredients with our proposed platform.  
An important extension would be to replace the
 Landau level quantization by Chern-band or anomalous Quantum Hall
platforms, where synthetic orbital structures could be engineered without a
large applied magnetic field.


\section*{Acknowledgments} 

This work has been supported in part by the Keck Foundation (Award No.~CRM:0132347) and by the Laboratory Directed Research and Development program and Sandia University Partnerships Network ( SUPN) program and performed in part at the Center for Integrated Nanotechnologies ( CINT), an Office of Science User Facility operated for the U.S. Department of Energy (DOE) Office of Science. Sandia National Laboratories is a multimission laboratory managed and operated by National Technology and Engineering Solutions of Sandia, LLC., a wholly owned subsidiary of Honeywell International, Inc., for the U.S. Department of Energy's National Nuclear Security Administration under Contract No. DE-NA0003525. This article describes objective technical results and analysis. Any subjective views or opinions that might be expressed in the article do not necessarily represent the views of the U.S. Department of Energy or the United States Government.

\appendix

\section{Estimate of LC-induced gap and Coulomb cost}

We describe in more detail our estimates used in the main text.  We will use the numerical values of the parameters as in the main text, namely, 
\[
\begin{gathered}
R_{\rm QH}=R_{\rm loop}=4~\mu{\rm m},\quad
B_0=1~{\rm T},\quad L=1~{\rm pH},\\
h_{\rm LC}=2~{\rm nm},\quad d=30~{\rm nm}.
\end{gathered}
\]
The magnetic length and Landau-level degeneracy are
\[
\ell_B=\sqrt{\frac{\hbar}{eB_0}}=25.7~{\rm nm},\quad
N_\phi=\frac{B_0\pi R_{\rm QH}^2}{\Phi_0}\simeq1.2\times10^4 .
\]
For an LC mode of frequency \(f_{\rm LC} = \omega_{\rm LC}/2\pi \),
\[
I_{\rm zpf}=\sqrt{\frac{\hbar\omega_{\rm LC}}{2L}},\quad
B_{\rm zpf}\sim \frac{\mu_0 I_{\rm zpf}}{2R_{\rm loop}} .
\]
The corresponding zero-point flux through the 2DEG sample is
\[
\frac{\Phi_{\rm zpf}}{\Phi_0}
\sim
\frac{B_{\rm zpf}\pi R_{\rm QH}^2}{\Phi_0}.
\]

A crude estimate of the orbital matrix element at the edge of the sample is obtained
from the paramagnetic vertex,
\[
\gamma_{\rm edge}
\sim
\frac{e}{m^\ast}A_{\rm zpf}(R)p_{B,\phi}(R)
\sim
\frac{e^2B_0B_{\rm zpf}R_{\rm QH}^2}{4m^\ast},
\]
where \(A_{\rm zpf}(R)\sim B_{\rm zpf}R/2\) and
\(p_{B,\phi}(R)\sim eB_0R/2\).  For a loop profile with the first harmonic
weight \(\epsilon_1=|\gamma_1/\gamma_0|\simeq0.3\),
\[
\gamma_1\simeq \epsilon_1\gamma_{\rm edge},\quad
V_{\rm LC}^{(1)}\sim \frac{\gamma_1^2}{\hbar\omega_{\rm LC}},\quad
\lambda_{\rm LC}^{\rm odd}\sim N_{\rm act}V_{\rm LC}^{(1)} ,
\]
with \(N_{\rm act}\sim10^3\) active LLL orbitals.  The resulting synthetic
Kitaev gap is estimated from
\[
E_{\rm gap}
=
\min_q
\sqrt{
\xi^2(q)+4|\Delta_1|^2\sin^2q
},
\]
where $\xi(q)=
-\mu_{\rm eff}-2t_1\cos q-2t_2\cos2q$.

Near \(\mu_{\rm eff}\simeq0\) and for \(|t_2|\ll |t_1|\),
\[
E_{\rm gap}\simeq 2|\Delta_1| .
\]

The numerical estimates are collected in Table~\ref{tab:scales}.  The GaAs values use \(m^\ast=0.067m_e\), \(\epsilon_r=12.9\), and
\(f_{\rm LC}=100~{\rm GHz}\); the 
InSb values are obtained using the same geometry but with
\(m^\ast=0.014m_e\), \(\epsilon_r=16.8\), and \(f_{\rm LC}=300~{\rm GHz}\).

\begin{table}[t]
\caption{
Representative LC-induced scales for GaAs and InSb quantum Hall droplets.
The pairing scale is quoted as an estimate because it depends on the
active-window profile and the self-consistent odd-channel susceptibility.
}
\label{tab:scales}
\begin{ruledtabular}
\begin{tabular}{lcc}
 & GaAs & InSb \\
\hline
\(f_{\rm LC}\) & \(100~{\rm GHz}\) & \(300~{\rm GHz}\) \\
\(\hbar\omega_{\rm LC}\) & \(414~\mu{\rm eV}\) & \(1.24~{\rm meV}\) \\
\(I_{\rm zpf}\) & \(5.8~\mu{\rm A}\) & \(10~\mu{\rm A}\) \\
\(B_{\rm zpf}\) & \(0.9~\mu{\rm T}\) & \(1.6~\mu{\rm T}\) \\
\(\Phi_{\rm zpf}/\Phi_0\) & \(1.1\times10^{-2}\) & \(1.9\times10^{-2}\) \\
\(\gamma_{\rm edge}\) & \(9.5~\mu{\rm eV}\) & \(79~\mu{\rm eV}\) \\
\(\gamma_1\) & \(2.9~\mu{\rm eV}\) & \(24~\mu{\rm eV}\) \\
\(V_{\rm LC}^{(1)}\) & \(2.0\times10^{-2}~\mu{\rm eV}\) & \(0.45~\mu{\rm eV}\) \\
\(\lambda_{\rm LC}^{\rm odd}\) & \(20~\mu{\rm eV}\) & \(0.45~{\rm meV}\) \\
\(\Delta_1\) & \(10-20~\mu{\rm eV}\) & \(0.2-0.45~{\rm meV}\) \\
\(E_{\rm gap}\) & \(20-40~\mu{\rm eV}\) & \(0.45-0.9~{\rm meV}\) \\
\(T_{\rm gap}\) & \(0.2-0.5~{\rm K}\) & \(5-10~{\rm K}\) \\
\(E_{\rm gap}/\hbar\omega_{\rm LC}\) & \(0.05-0.10\) & \(0.36-0.72\) \\
\end{tabular}
\end{ruledtabular}
\end{table}

We next estimate the screened Coulomb cost in the same adjacent-orbital
pairing channel.  For a metallic screening plane at distance \(d\),
\[
V_C(q)
=
\frac{e^2}{2\epsilon_0\epsilon_r q}
\left(1-e^{-2qd}\right).
\]
The LLL pseudopotentials are
\[
V_\ell^C(d)
=
\frac{e^2}{4\pi\epsilon_0\epsilon_r\ell_B}
\int_0^\infty dx\,
\left(1-e^{-2dx/\ell_B}\right)
L_\ell(x^2)e^{-x^2}.
\]
The antisymmetrized adjacent-orbital cost is
\[
U_m^C(d)
=
\langle m,m+1|V_C|m,m+1\rangle_A
=
\sum_{k=0}^{m}
w_{m,k}V_{2k+1}^C(d),
\]
with
\[
w_{m,k}
=
\frac{
\binom{m}{k}^2
(2m-2k)!\,(2k+1)!
}{
4^m\,m!\,(m+1)!
}.
\]
For \(m=0\), \(U_0^C=V_1^C\), which yields 
\[
U_0^C\simeq0.67~{\rm meV}\quad{\rm(GaAs)},\quad
U_0^C\simeq0.51~{\rm meV}\quad{\rm(InSb)} .
\]
Thus, pairing near the origin $R = 0$ is electrostatically unfavorable.
This constraint is removed in an annulus of a finite radius.  For
\[
R_{\rm in}=2~\mu{\rm m},\quad R_{\rm out}=4~\mu{\rm m},
\]
the active orbital window is
\[
m_{\rm in/out}\simeq \frac{R_{\rm in/out}^2}{2\ell_B^2},
\quad
m_{\rm in}\simeq3.0\times10^3,\quad
m_{\rm out}\simeq1.2\times10^4 .
\]
In this large-\(m\) window the screened adjacent-bond repulsion is strongly
suppressed:
\begin{align*}
{\rm GaAs:}\;&
U^C_{m_{\rm in}}\sim2\times10^{-2}~\mu{\rm eV},
\;
U^C_{m_{\rm out}}\sim3\times10^{-3}~\mu{\rm eV},\\
{\rm InSb:}\;&
U^C_{m_{\rm in}}\sim1.5\times10^{-2}~\mu{\rm eV},
\;
U^C_{m_{\rm out}}\sim2.3\times10^{-3}~\mu{\rm eV}.
\end{align*}
These values are negligible compared with the estimated LC-induced gaps in
Table~\ref{tab:scales}.  The annular geometry therefore removes the
small-\(m\) orbitals where Coulomb repulsion is largest and places the
synthetic Kitaev chain in a large-radius window where screened
electrostatic repulsion does not compete with the induced odd-channel
pairing.

Finally, the relevant single-particle gaps remain favorable.  At
\(B_0=1~{\rm T}\),
\[
\hbar\omega_c
=
\frac{\hbar eB_0}{m^\ast}
\simeq
1.73~{\rm meV}\quad{\rm(GaAs)},\qquad
8.3~{\rm meV}\quad{\rm(InSb)} .
\]
Thus the LC mode lies below the inter-Landau-level transition in both
materials.  In InSb, the Zeeman scale is also large:
\[
E_Z=|g^\ast|\mu_BB_0\simeq2.0~{\rm meV}
\qquad
(|g^\ast|\simeq35),
\]
supporting a spin-polarized description.  The main constraint in the InSb
case is instead a marginal dispersive hierarchy,
\(E_{\rm gap}/\hbar\omega_{\rm LC}\sim0.36-0.72\); a conservative regime
therefore uses the lower part of the estimated gap range or a smaller
active window \(N_{\rm act}\).

\bibliographystyle{apsrev4-2}
\bibliography{References}

\end{document}